\documentclass[twocolumn,showpacs,preprintnumbers,amsmath,amssymb]{revtex4-2}

\usepackage{bm}

\usepackage{hyperref}
\usepackage{graphicx}
\usepackage{mwe}
\usepackage{subfigure}
\usepackage{dcolumn}
\usepackage{xcolor}

\usepackage{float}
\usepackage[capitalize]{cleveref}
\hypersetup{
    colorlinks=true,
    linkcolor=blue,
    citecolor=blue, allcolors=blue
}
\usepackage{figsize}
\usepackage{mathtools}

\begin{document}
\title{Klein Tunneling and Fabry-Pérot Resonances in Twisted Bilayer Graphene}
\author{A. Bahlaoui$^{1}$ and Y. Zahidi$^{1}$}
\affiliation{$^{1}$\em Multidisciplinary Research and Innovation Laboratory, EMAFI team, Polydisciplinary Faculty, Sultan Moulay Slimane University, 25000 Khouribga, Morocco}

\begin{abstract}
	
The paper discusses the Klein tunneling and Fabry-Pérot resonances of charge carriers through a rectangular potential barrier in twisted bilayer graphene. Within the framework of the low-energy excitations, the transmission probability and the conductance are obtained depending on the parameters of the problem. Owing to the different chirality in twisted bilayer graphene, the propagation of charge carriers exhibits an anisotropic behavior in transmission probability and Fabry-Pérot resonances. Moreover, we show that the anisotropy of the charge carriers induces asymmetry and deflection in the Fabry-Pérot resonances and Klein tunneling, and they are extremely sensitive to the height of the potential applied. Additionally, we found that the conductance is strongly sensitive to the barrier height but weakly sensitive to the barrier width. Therefore, it is possible to control the maxima and minima of the conductance of charge carriers in twisted bilayer graphene. With our results, we gain an in-depth understanding of tunneling properties in twisted bilayer graphene, which may help in the development and designing of novel electronic nanodevices based on anisotropic 2D materials.


\end{abstract}
\maketitle
\section{INTRODUCTION}

The discovery of massless relativistic Dirac fermions in graphene has revolutionized condensed matter physics, revealing a variety of exotic electronic properties with promising applications \citep{novoselov2005two,neto2009electronic}. Among these, the Klein paradox, a quantum tunneling phenomenon originally predicted in relativistic quantum mechanics, which has been experimentally realized in graphene \citep{katsnelson2006chiral}. In this system, charge carriers incident normally on an electrostatic potential barrier experience perfect transmission, regardless of the barrier’s height and width. This remarkable effect arises from the chiral nature of quasiparticles in graphene and has been extensively studied in analogy with Klein’s gedanken experiment \citep{young2009quantum}. However, for oblique incidence, the transmission probability exhibits interference patterns akin to Fabry-Pérot resonances in optical cavities \citep{PhysRevLett.101.156804}, a behavior that has been observed experimentally in \textit{npn} junctions \citep{Velasco_2009}.

In bilayer graphene, the nature of charge carriers fundamentally changes due to the quadratic dispersion relation and the different chirality of quasiparticles \citep{peres2006electronic}. Unlike in monolayer graphene, which exhibits perfect transmission at normal incidence, bilayer graphene exhibits perfect reflection at normal incidence, highlighting the crucial role of chirality in quantum tunneling \citep{katsnelson2006chiral}. This difference in transmission behavior between monolayer and bilayer graphene suggests the possibility of engineering electronic transport properties by controlling chirality. However, in twisted bilayer graphene (TBG), where two graphene layers are stacked with a relative twist angle, the electronic structure becomes more complex due to moiré-induced modifications of the band structure \citep{he2013chiral}. These modifications introduce additional degrees of freedom that allow for tunable chiral tunneling, where the transmission probability can be continuously tuned from perfect transmission to partial or complete reflection by adjusting the barrier parameters .

In addition, the chirality of quasiparticles in TBG introduces an extra dimension of anisotropy to their electronic and optical behaviors. In contrast to monolayer graphene, where group velocity and momentum are collinear, TBG displays non-collinear group velocity, leading to significant anisotropic transport effects \citep{Zhang_2019,park2008anisotropic}. This anisotropy is evident in quantum tunneling phenomena, such as the directional dependence of Klein tunneling and Fabry-Pérot resonances, in which charge carriers are deflected rather than passing straight through barriers \citep{PhysRevB.97.235113}. These unique and unconventional transport properties offer deeper insights into the relationship between moiré physics and chiral tunneling mechanisms.

The fact that the tunneling behavior can be tuned and controlled in TBG opens new opportunities for innovative applications in electron optics and valleytronics. The interaction between moiré band engineering and chiral tunneling may help in the development and design of adjustable electron lenses, valley filters, and quantum transport devices \citep{he2013chiral,BAHLAOUI2024115880,ren2024electron,stauber2018chiral,duarte2024moire}. Gaining a thorough understanding of these effects is crucial for advancing quantum transport theories in moiré-engineered materials and unlocking the full potential of TBG for future nanoelectronic technologies.

In this study, we investigate the anisotropic tunnel
ing behavior of TBG, with a partic
ular focus on the manifestation of Klein tunneling and
Fabry-Pérot resonances in this system. To achieve this goal, we have organized the rest of this paper as follows. In Section \ref{tsection2}, we first use  an effective two-band model to describe the electronic properties of TBG. We then derive the wave functions of charge carriers penetrating through a rectangular potential barrier. Next, the tunneling problem is solved straightforwardly to compute the transmission probability and conductance. Section \ref{tsection3} presents a systematic analysis of numerical results for normal and non-normal incidence, followed by a discussion of our findings and their broader implications in Section \ref{tsection4}.

\section{MODEL}\label{tsection2}

\begin{figure}[!b]
	\centering
 \includegraphics[scale=.25]{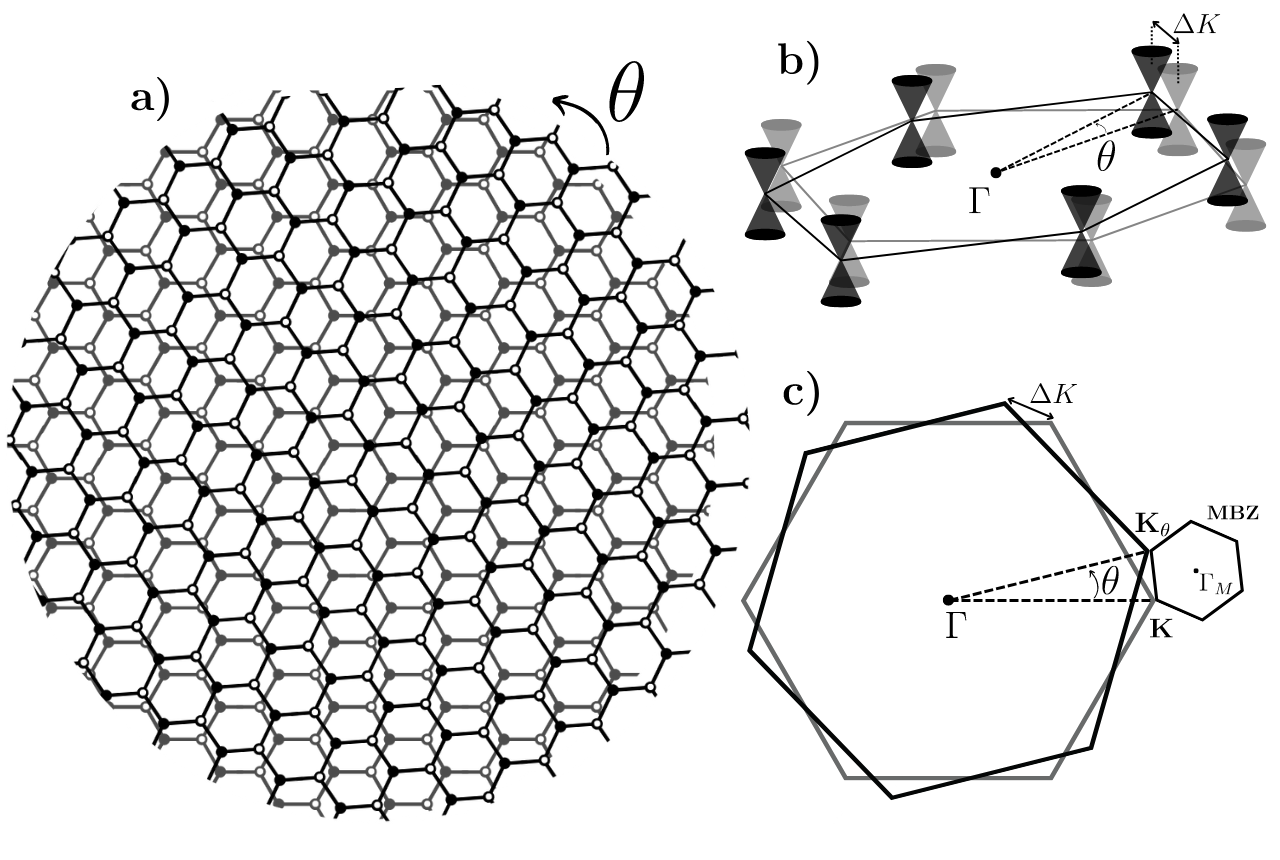} 	
	{\tiny \caption{\label{fig:1TBG} Structural diagram of TBG. \textbf{(a)} A schematic representation of TBG can be obtained by the relative rotation of the upper and lower layers to each other by a twist angle of $\theta$. \textbf{(b)} The schematic shows how the relative rotation separates the Dirac cones of the two graphene layers of the TBG in reciprocal space, resulting in a relative shift $\Delta K$ between the corresponding Dirac points of the TBG. \textbf{(c)} The mini-Brillouin zone mBZ of the Moiré superlattice of TBG is generated from a relative shift $\Delta K$ between the wavefunctions of Dirac points $K$ and $K_{\theta}$.}}
\end{figure}

We consider a TBG, which can be viewed as a relative rotation to each other of the upper-lower layers in Bernal stacked bilayer graphene, by a twist angle $\theta$ around a common vertical axis, as shown in \cref{fig:1TBG} {\color{blue}(a)}. This rotation leads to a displacement and hybridization between Dirac cones in the mini-Brillouin zone (mBZ). This redistribution of the electronic band structure in twisted bilayer graphene results in the appearance of two saddle points, $K$ and $K_{\theta}$, which no longer coincide \cite{dos2007graphene,dos2012continuum}, as shown in \cref{fig:1TBG}{\color{blue}(b)} and \cref{fig:1TBG}{\color{blue}(c)}. In addition, the zero energy states occur at $\vec{k}$=$-\left(\Delta K_{x} / 2,\Delta K_{y}/2\right)$ and $\vec{k}$=$\left(\Delta K_{x}/2,\Delta K_{y}/2\right)$ in layers 1 and 2, respectively. Where $\left(\Delta K_{x},\Delta K_{y}\right)$ is  the relative shift of Dirac points in the two layers, and its modulus is defined as $\Delta K$=$2\lvert K\rvert\sin(\theta/2)$, where $\lvert K\rvert$=$4\pi/3a_{0}$ is the momentum-space dependence via the graphene lattice constants in real-space $a_{0}=\sqrt{3}a\approx0.25\mathrm{~nm}$.

At low energies, twisted bilayer graphene can be effectively described using the continuum nearest-neighbor tight-binding model \cite{partoens2007normal,neto2009electronic}. Thus, the effective Hamiltonian can be written as \cite{he2013chiral}
\begin{small}
\begin{equation}
\label{eqn:h1}
H^{eff}=-\frac{2 v_{F}^{2}}{15\tilde{t}_{\perp}}\left[\begin{array}{cc}
0 & \left(k^{\dagger}\right)^{2}-\left(k_{\theta}^{\dagger}\right)^{2} \\
\left(k\right)^{2}-\left(k_{\theta}\right)^{2}  & 0
\end{array}\right],       
\end{equation}
\end{small}
where $v_{\mathrm{F}}$\;=\;$10^{6}\mathrm{~m/s}$ denotes the Fermi velocity. $t_{\perp}$\!\;$\approx$\;0.27eV is the interlayer coupling, can be well approximated for a small twist angle by $\tilde{t}_{\perp}$\!\;$\simeq$ \!0.4$t_{\perp}$  \cite{dos2012continuum}. $k=\left(k_{x}+i k_{y}\right)$ and $k^{\dagger} = \left(k_{x}-i k_{y}\right)$ are the in-plane wave vector and its conjugate, where $k_{x,y}$ =-$i\partial_{x,y}$. In the presence of twist angle between layers ($\theta$\!\;$\neq$\;0), the complex wave vector in mBZ and its conjugate are defined as $k_{\theta}$=$\Delta K/2$=$\left(\Delta K_{x}+i \Delta K_{\mathrm{y}}\right)/2$, and $k_{\theta}^{\dagger}$=$\Delta K^{\dagger}/2$=$\left(\Delta K_{x}-i \Delta K_{y}\right)/2$, respectively. For simplicity, we later set $\Delta K_{x}=0$ and $\Delta K_{y}=\Delta K$. However, this continuum model is valid only for small twist angles $a\!\ll\!L$ (here $L$ is period of the moire patterns in TBG) \cite{he2013chiral}. Indeed, the  Van Hove singularities (VHSs) in a TBG have been detected experimentally for $\theta$\!\;$\leqslant$\!\;$10^{\circ}$ \cite{ohta2012evidence,li2010observation,luican2011single,PhysRevLett.109.126801,PhysRevLett.109.196802,PhysRevB.85.235453,PhysRevLett.108.246103}. Therefore, for $\theta$\!\;$\leqslant\!10^{\circ}$, the Hamiltonian (\ref{eqn:h1}) remains a valid approximation, with $\left(L\sim1.4\mathrm{~nm}\right.$ at $\left.\theta=10^{\circ}\right)$. With the effective Hamiltonian established, we now turn our attention to studying the transport properties of TBG by analyzing the quantum tunneling of quasiparticles through a potential barrier.\\

\begin{figure}
\centering
 \includegraphics[scale=.4]{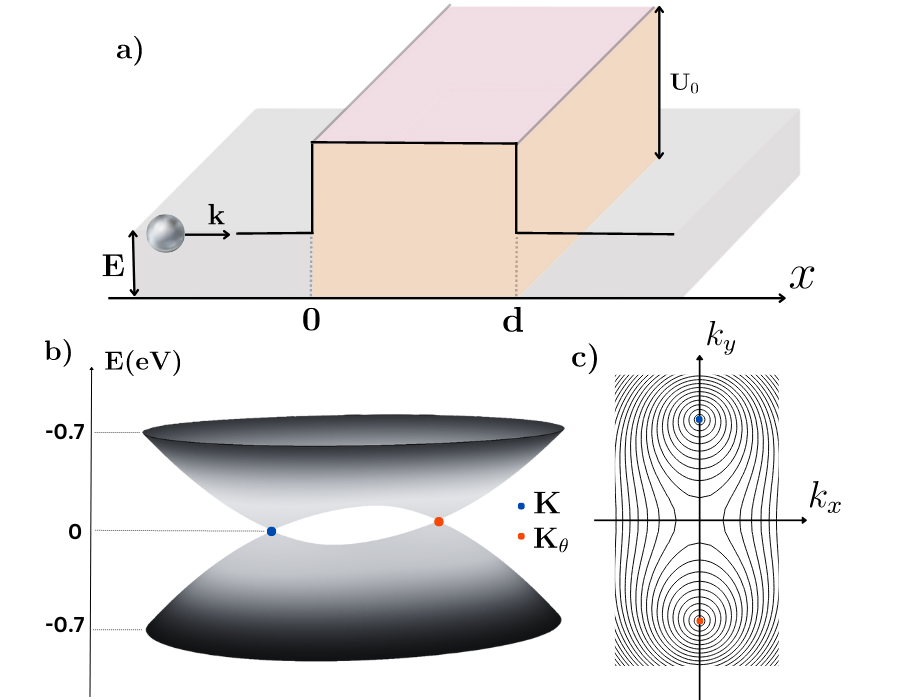} 
\caption{Quantum tunneling through a single barrier in TBG.  \textbf{(a)} Schematic diagram of a quasiparticle coming with a incident energy $E$ to a potential barrier of height $U_{0}$ and width $d$. We assume that the single barrier has a rectangular shape and is infinite along the y-direction. \textbf{(b)} The low-energy spectrum of TBG with a twist angle of $\theta=3.89^{\circ}$. The separation of the Dirac cones forms two saddle points, $K$ and $K_{\theta}$. \textbf{(b)} A contour plot of the energy spectrum of TBG around $K$ and $K_{\theta}$. $k_{x}$ have a perpendicular direction to the barrier.}
\label{fig00}
\end{figure}

To study resonant tunneling in TBG, we analyze the scattering of quasiparticles with incident energy $E$ through a single barrier structure $U_{0}(x)$, which has a thickness $d$ and extends infinitely along the $y$-axis. The system consists of three distinct regions: the left region $\left(x<0\right)$, the barrier region $\left(0<x<d\right)$, and the right region $\left(x>d\right)$, as illustrated in \cref{fig00}{\color{blue}(a)}. Unlike in pristine graphene systems, the group velocity $\left(\vec{\nu}_{k}=(1 / \hbar)\left(\nabla_{k} E\right)_{k}\right)$ in twisted bilayer graphene is not anymore parallel to quasi-momentum $k$. Therefore, we assume a rectangular potential barrier, which prevents intervalley scattering between $K$ and $K_{\theta}$ \citep{katsnelson2006chiral,he2013chiral}. Thus, we analyze the scattering process within a single valley, neglecting intervalley coupling.\\
 
The wave function of TBG is given by the vector {\footnotesize $\Phi(x, y)$=$\left(\begin{array}{c}\psi^{A}(x), \psi^{B}(x)\end{array}\right)^{T} e^{i k_{y} y}$}. Substituting this wave function into the equation $H^{eff} \Phi=E \Phi$, we obtain the energy spectrum derived from the Hamiltonian (\ref{eqn:h1}) as 
\begin{small}
\begin{equation}
\label{eqn:h2}
E^{2}=\left(\frac{2 v_{F}^{2}}{15 \tilde{t}_{\perp}}\right)^{2}\left[\left(k_{x}^{2}-k_{y}^{2}+\left(\frac{\Delta K}{2}\right)^{2} \right)^{2}+\left(2 k_{x} k_{y}\right)^{2}\right].
\end{equation}
 \end{small} 
\Cref{fig00}{\color{blue}(b)} displays the electronic band structure of the quasiparticles in close proximity to one of the two valleys in TBG with a finite interlayer coupling. Thus, in our analysis, we focus on nearest-neighbor interlayer hopping in the Hamiltonian (\ref{eqn:h1}). The energy dispersion consists of two symmetrical valley bands characteristic of chiral massless fermions, corresponding to electron-like and hole-like states. This symmetry plays a crucial role in chiral tunneling, which is observed in various graphene systems, including monolayer graphene, Bernal stacked bilayer graphene, and TBG \citep{katsnelson2006chiral}. However, when large next-nearest-neighbor interlayer hopping is considered, electron-hole symmetry is broken, leading to the suppression of chiral tunneling \citep{PhysRevLett.61.2015}. The saddle points in the electronic band structure manifest as VHSs in the density of states (DOS), leading to two low-energy VHSs for low-energy at $\pm E_{V}$=$\pm 1 / 2\left(\hbar v_{\mathrm{F}}|\Delta K|-2 t_{\perp}\right)$ (\cref{fig00}{\color{blue}(b)} and \cref{fig00}{\color{blue}(c)}) \cite{cao2018correlated,dos2007graphene,dos2012continuum}. More experimental and theoretical results have shown that the VHS position of TBG depends on the twisted angle $\theta$ \cite{PhysRevLett.109.126801,PhysRevLett.109.196802,li2010observation}. These features have been experimentally validated using Raman spectroscopy, scanning tunneling spectroscopy, and angle-resolved photoemission spectroscopy \cite{ohta2012evidence,li2010observation,luican2011single,PhysRevLett.109.126801,PhysRevLett.109.196802,PhysRevB.85.235453,PhysRevLett.108.246103}. In Bernal stacked bilayer graphene, electron-electron interactions can lead to a splitting of the quadratic band-touching point into two Dirac points, resulting in an energy dispersion similar to that of TBG. This phenomenon has been associated with a nematic symmetry-breaking state \cite{PhysRevB.81.041401,PhysRevLett.108.186804,PhysRevB.82.201408}.\\\\
\begin{figure*}[!t]
\centering
\includegraphics[scale=.9]{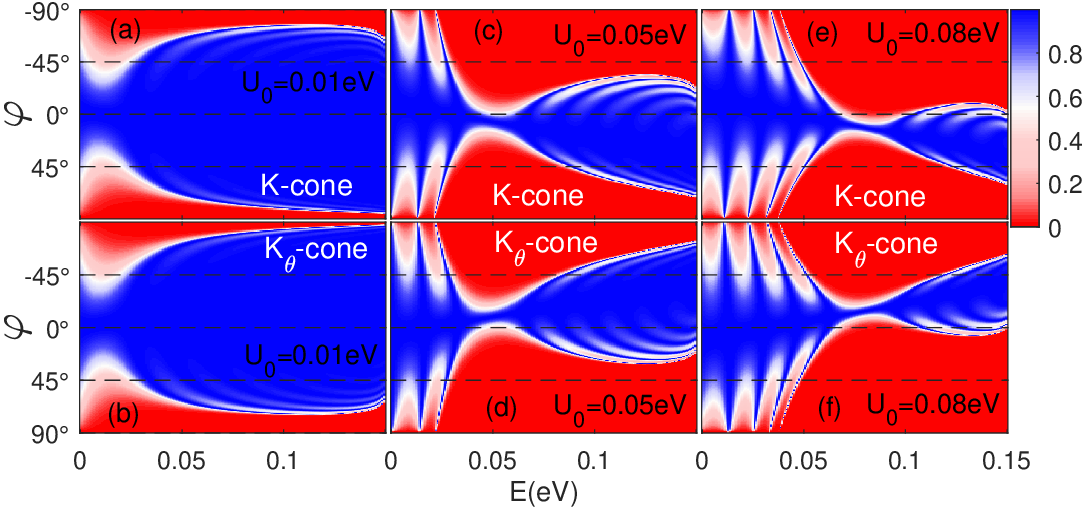}  \caption{Contour plot of the transmission probability of quasiparticles in $K$-cone : (a), (c), (e), and $K_{\theta}$-cone : (b), (d), (f), as a function of the energy $\left(E\right)$ and of the incident angle $\left(\varphi\right)$ for barrier
height values $U_{0}=0.01eV$, $U_{0}=0.05eV$ and $U_{0}=0.08eV$. The remaining parameters are the twist angle $\theta=3.89^{\circ}$, and barrier width $d=100nm$.}
\label{figT0}
\end{figure*}

By substituting the full spatial part of a trial wave function  $\Phi(x, y)$ into the eigenvalue equation $H \Phi=E \Phi$, with the Hamiltonian given in (\ref{eqn:h1}), we obtain the eigenstates in the three regions. This allows us to compute the transmission probability of a quasiparticle traversing a single potential barrier and explore resonance tunneling effects in TBG by varying system parameters. Transport properties can then be evaluated using the transmission probability to compute the conductance function via the Landauer–Büttiker formalism \citep{BLANTER20001}
\begin{equation}
G(E)= \frac{G_0}{2 \pi} \int_{-\infty}^{\infty} d k_y \sum_{l, m= \pm} T_m^l\left(E, k_y\right).
\end{equation}
Here, $G_0=4 e^2 W / h$ represents the fundamental conductance factor, where $e$ is the elementary charge, $W$ is the sample width along the $y$-direction, and $h$ is Planck’s constant.

\section{Results and discussion}\label{tsection3}

In this section, we analyze the dependence of transmission and conductance on the incident angle ($\varphi$) and structural parameters $\left(E, U_{0}, d\right)$ for quasiparticles in both the $K$ and $K_{\theta}$ valleys.

In \cref{figT0}{\color{blue}(a)(c)(e)} and \cref{figT0}{\color{blue}(b)(d)(f)}, we present the $\left(E, \varphi\right)$ maps of the transmission probabilities of quasiparticles in the $K$ and $K_{\theta}$ cones, respectively. The transmission profiles for quasiparticles in the two cones are asymmetric about normal incidence ($\varphi=0^{\circ}$), and this asymmetry intensifies with increasing potential barrier height and incident energy \citep{he2013chiral, PhysRevB.90.075410, BAHLAOUI2024115880}.
Unlike the tunneling process of quasiparticles in monolayer graphene (massless Dirac fermions) and Bernal graphene (massive chiral fermions), the quasiparticles in the $K$ and $K_{\theta}$ cones of TBG are anisotropic massless Dirac fermions, which accounts for the observed asymmetry in the tunneling profiles. However, the transmission probabilities for quasiparticles in the $K$ and $K_{\theta}$ cones exhibit mirror-symmetric behavior about the incident angle $\varphi=0^{\circ}$. This mirror symmetry implies that the transmission probabilities follow the relation: $T_{K}\left(\varphi \right)  = T_{K_{\theta}}\left(-\varphi \right)$ and $T_{K}\left(-\varphi \right) = T_{K_{\theta}}\left(\varphi \right)$. However, this symmetry depends on the barrier orientation and may break down if the barrier is not parallel to the line connecting $K$ and $K_{\theta}$ \citep{he2013chiral}.
Without the twist effect $\left(\theta=0^{\circ}\right)$ \citep{allain2011klein}, the transmission profiles exhibit two well-known phenomena: ${T}_{K, K_{\theta}}=1$ at normal incidence $\left(\varphi=0^{\circ}\right)$, independent of the barrier height $U_{0}$ (i.e., Klein tunneling) and resonant peaks in oblique directions (i.e., Fabry-Pérot resonances). For a small value of $U_{0}$, when electrons are incident normally on the potential barrier, the transmission probability for the two cones is perfect and does not change with the incident energy (i.e., Klein tunneling \citep{katsnelson2006chiral}), as shown in \cref{figT0}{\color{blue}(a)} and \cref{figT0}{\color{blue}(b)}. This behavior can even occur in the vicinity of the normal incidence. Additionally, we show that the Fabry-Perot resonance peaks are not visible in the transmission profiles in $K$ and $K_{\theta}$ at small barrier height $U_{0}$. However, as $U_{0}$ increases for $K$-cone, as illustrated in \cref{figT0}{\color{blue}(c),(e)}, and $K_{\theta}$-cone, as illustrated in \cref{figT0}{\color{blue}(d),(f)}, transmission resonance peaks appear in oblique directions (i.e., for non-zero incident angles), particularly at certain Fermi energies in the energy region $E<U_{0}$ \citep{katsnelson2006chiral}. It is clearly seen that the number of these resonance peaks increases with the potential barrier height, and the energy positions of the tunneling resonances shift as the incident angle changes. Thus, Fabry-Perot resonances are highly sensitive to the potential barrier height because when the barrier height $U_{0}$ is larger than the incident energy $E$, additional resonant tunneling peaks appear from the propagating of the incident electrons into the valence band. This is a characteristic of the Klein tunneling \citep{katsnelson2006chiral,beenakker2008colloquium}. In TBG, unlike untwisted graphene systems, the twist effect mainly influences the symmetry of the Fabry-Pérot resonance behavior. Therefore, we discuss how the asymmetry and deflection of the Fabry-Pérot resonances affect the Klein tunneling behavior of quasiparticles in TBG. Regarding the mirror-symmetric behavior of the two cones mentioned earlier, in what follows, we consider only the scattering electrons from one valley, $K$.\\ 

\begin{figure}[!b]
\centering
\includegraphics[scale=.75]{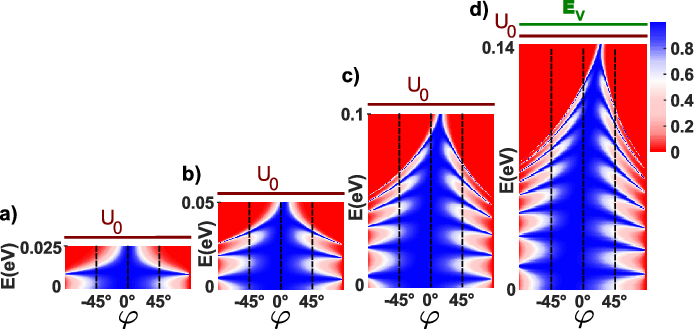}    
\caption{Contour map of the transmission probability of quasiparticles in $K$-cone ${T}_{K}\left(E,\varphi\right)$, at small Fermi energy $\left(E<E_{V}\right)$. We consider the energy region $E<U_{0}$ for barrier height values (a) $U_{0}=0.03eV$, (b) $U_{0}=0.055eV$,(c) $U_{0}=0.105eV$, and (d) $U_{0}=0.145eV$. The remaining parameters are the twist angle $\theta=3.89^{\circ}$, barrier width $d=100nm$, and $E_{V}=0.15eV$.}
\label{figT044}
\end{figure}

The contour map for transmission probability ${T}_{K}\left(E,\varphi\right)$ of quasiparticles in $K$-cone at Fermi energy $\left(E<E_{V}\right)$ for different barrier height $U_{0}$, is shown in \cref{figT044}. The contour plots reveal an asymmetry in the Fabry-Pérot resonances about normal incidence. As the barrier height $U_{0}$ increases, the Fabry-Pérot resonance becomes more asymmetric.  For small values of $U_{0}$, the Klein tunneling always holds regardless of the barrier height and the incidence energy value (as illustrated in \cref{figT044}{\color{blue}(a)}). As $U_{0}$ increases, perfect transmission shifts away from normal incidence, particularly when the incident energy approaches the barrier height (as illustrated in \cref{figT044}{\color{blue}(b)}). This deflection intensifies as $U_0$ increases, further enhancing the asymmetry of Fabry-Pérot resonances, as shown in \cref{figT044}{\color{blue}(c)-(d)}. The asymmetry and deflection of Fabry-Pérot resonances in TBGe arise primarily from the moiré pattern and the resulting anisotropic energy dispersion \citep{PhysRevB.97.235113}. This remarkable character significantly influences the electronic transport of Dirac fermions in TBG; the Klein tunneling under the twisting effect is still preserved but can be achieved with two separated oblique directions for the two Dirac fermions of opposite chiralities. Note additionally that, unlike the case of monolayer or Bernal bilayer graphene, the group velocity $\vec{v}$ of wave packets in TBG is non-collinear to its wave vector $\vec{k}$ \textcolor{blue}{\citep{he2013chiral}}. This is one of the key elements that can also explain the deflection of Klein tunneling and Fabry-Pérot resonances.\\
\begin{figure}[!t]
\centering
\includegraphics[scale=.7]{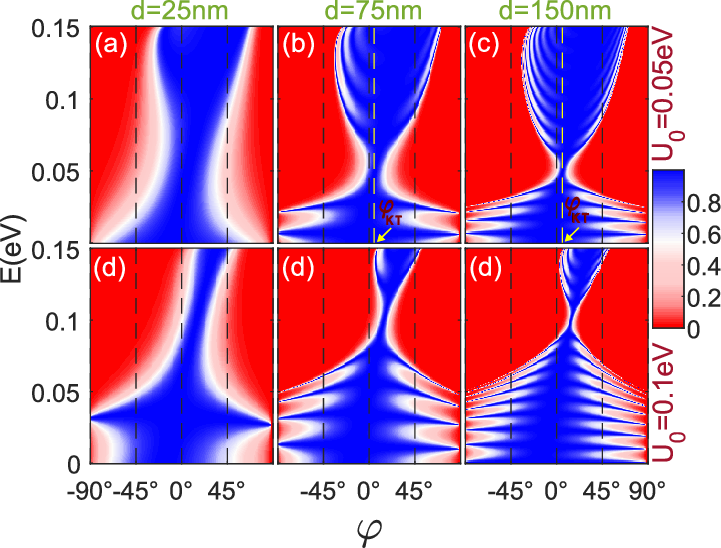}  \caption{(color online) contour plots of the transmission probabilities of quasiparticles at the $K$ cone for different barrier width values (a) $d=25nm$, (b) $d=50nm$, (c) $d=150nm$ et (d) $d=200nm$. The remaining parameters are the twist angle $\theta=3.89^{\circ}$, and barrier height $U_{0}=0.05eV$.}
\label{figT04}
\end{figure}
Next, it is interesting to see the effect of the barrier width $\left(d\right)$ on Klein tunneling and Fabry-Perot resonances through the potential barrier. In \cref{figT04}{\color{blue}} we present $\left(E,\varphi\right)$ contour plots of the transmission probabilities of quasiparticles at the $K$-cone for different barrier width values $d=25nm$, $d=75nm$ and $d=150nm$, with barrier height of $U_{0}=0.05eV$ (upper panels) and $U_{0}=0.1eV$ (lower panels). 

In general, transmission resonances become more prominent as the barrier width increases. Also, the numbers and positions of resonant peaks are remarkably changed with the increase in barrier width. We also find an asymmetric behavior of the transmission with respect to the normal incidence $\left(\varphi=0^{\circ}\right)$. In the upper panels of \cref{figT04}{\color{blue}}, we can see a perfect transmission around the normal incidence $\varphi=0^{\circ}$ (as illustrated in \cref{figT04}{\color{blue}(a)}) and around the non-normal incidence $\varphi\neq0^{\circ}$ (as illustrated in \cref{figT04}{\color{blue}(b)-(c)}), regardless of the incident energy $\left(E\right)$, and the potential barrier becomes fully transparent (i.e., Klein tunneling). An increased barrier width shifts Klein tunneling away from normal incidence and enhances the asymmetry of Fabry-Pérot resonances. We also observe that quasiparticles at the $K$-cone have non-resonant tunneling when the incident angle is $\varphi=0^{\circ}$ for $d=25nm$ (as shown in \cref{figT04}{\color{blue}(a)}) and $\varphi=\varphi_{KT}$ for $d=75nm,150nm$ (as shown in \cref{figT04}{\color{blue}(b)-(c)}), we note $\varphi_{KT}$ is the angle of the shifted Klein tunneling. In the case of the quasiparticles at the $K_{\theta}$-cone, we can predict that the Klein tunneling is shifted to the angle of incidence $\varphi=-\varphi_{KT}$, with non-resonant tunneling $-\varphi_{KT}$. Electrons at the $K$-cone ($K_{\theta}$-cone) with $\varphi=\varphi_{KT}$ ($\varphi=-\varphi_{KT}$) tunnel perfectly without trajectory deflection, regardless of the tunneling regime. Thus, varying the barrier width does not alter the electron trajectory in either cone. 

As $d$ increases, Klein tunneling shifts, and perfect transmission becomes less likely, especially when the incident energy approaches $U_{0}$. 
This is because the angular variation of the pseudospin changes, and due to the non-collinearity between the group velocity and pseudospin in anisotropic systems like TBG, Klein tunneling occurs along directions that are not necessarily normal to the barrier \citep{PhysRevB.98.205421}. In contrast, this behavior is significantly different when we increase the barrier height, as shown in the lower panels of \cref{figT04}{\color{blue}}. We see clearly that the deflection of the transmitted electrons is sensitive to the barrier height $U_{0}$ but it insensitive to an increase of barrier width $d$. 

\begin{figure}[!t]
\centering
\includegraphics[scale=.7]{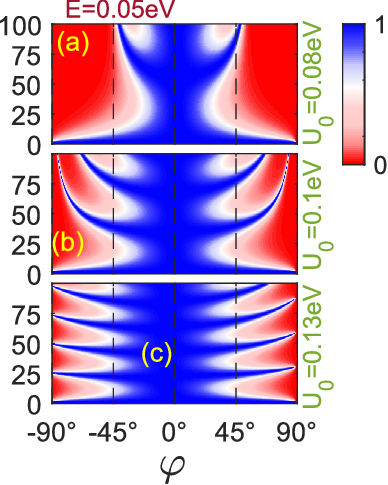}  \caption{$\left(d,\varphi\right)$ contour plots of the transmission probabilities of quasiparticles at the $K$-cone for different barrier height values (a) $U_{0}=0.08eV$, (b) $U_{0}=0.1eV$, et (c) $U_{0}=0.13eV$. The remaining parameters are the twist angle $\theta=3.89^{\circ}$, and incident energy is chosen fixed at $E=0.05eV$.}
\label{figT0666}
\end{figure}

In \cref{figT0666}, we present $\left(d,\varphi\right)$ maps of transmission probabilities of quasiparticles in the $K$-cone for different barrier height values at $E=0.08eV$. Notice that for small incidence angles $\varphi$, the transmission is perfect regardless of the magnitude of the barrier width $d$. For larger $\varphi$, the transmission probability oscillates as a function of $d$ with an oscillation period change as the barrier height changes.\\
\begin{figure}[!b]
\centering
\includegraphics[scale=.55]{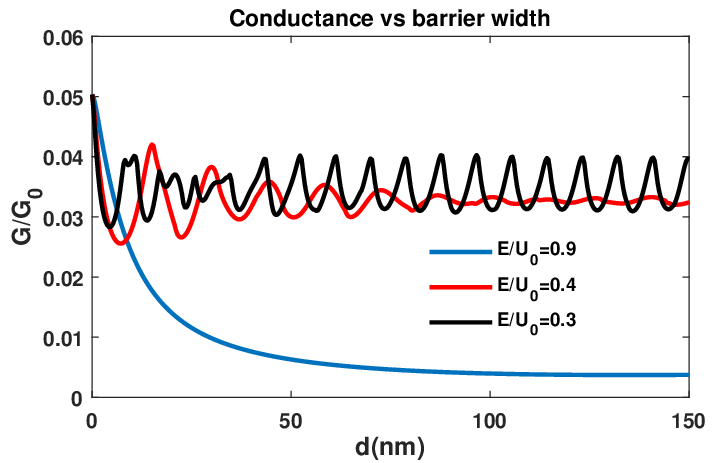}  
 \caption{(Color online) Conductance $G/G_{0}$ versus the width of the potential barrier $d$ under different ratios $E/U_{0}=0.3,\;0.4$ and $0.9$. The remaining parameters are the twist angle $\theta=3.89^{\circ}$, and incident energy is chosen fixed at $\left(E=0.08eV\right)$.}
\label{figT05}
\end{figure}
Next, we examine the transmission properties discussed above through measurable quantities such as conductance $G$ \citep{PhysRevLett.96.246802}. We now analyze the numerical results on tunneling conductance, which is easier to measure experimentally than the transmission coefficient. Therefore, we consider various ratios of $E/U_{0}$ that signifies the relationship between the incident energy $\left(E\right)$ and the barrier height $\left(U_{0}\right)$. 

We fix the incident energy at $E = 0.08$ eV and present the conductance $G/G_{0}$ as a function of barrier width $d$ for different values of $E/U_{0}$ in \cref{figT05}. As $d$ increases, the barrier reduces the total number of electrons transmitted, thereby decreasing the overall conductance. For small values of $E/U_{0}$, the conductance exhibits oscillations as a function of barrier width. However, as $E/U_{0}$ increases, these oscillations disappear for larger barrier widths and are particularly suppressed when the incident energy approaches the barrier height ($E/U_{0} = 0.9$). The oscillatory pattern in conductance reflects Fabry-Pérot resonances observed in tunneling transport. Incident electrons can create interference effects by undergoing multiple reflections within the potential barrier, leading to constructive or destructive interference that can enhance or suppress transmission, respectively \citep{PhysRevLett.121.127706}.\\
\begin{figure}[!t]
\centering
\includegraphics[scale=.6]{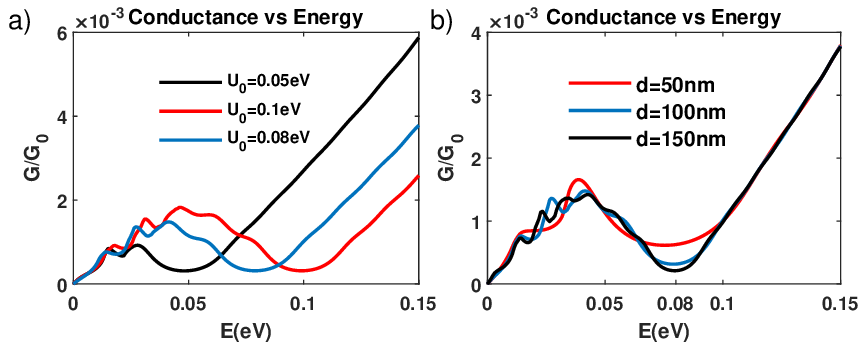} 
 \caption{(Color online) Conductance $G/G_{0}$ versus the incident energy $E$ under different values of the barrier (a) height and (b) width. The remaining parameters are the twist angle $\theta=3.89^{\circ}$, the barrier width in (a) chosen at $d=100nm$, and incident energy in (b) chosen fixed at $\left(U_{0}=0.08eV\right)$.}
\label{figT06}
\end{figure}

Furthermore, we present the dependence of conductance on incident energy for different values of the barrier height and width, as illustrated in \cref{figT06}{\color{blue}(a)} and \cref{figT06}{\color{blue}(b)}. Generally speaking, the conductance starts increasing from zero incident energy $E$ because no propagating modes exist at this energy; the density of states vanishes outside the potential barrier. The conductance is strongly sensitive to the barrier height $U_{0}$ but only weakly sensitive to the barrier width $d$. 

As $E$ increases, the conductance at a given $U_{0}$ rises to a maximum at half the barrier height ($E = U_{0}/2$) and then decreases to a nonzero minimum when the incident energy equals the barrier height (i.e., the density of states vanishes inside the potential barrier \citep{10.1063/1.5092512}), as shown in \cref{figT06}{\color{blue}(a)}. The minimum of the conductance is independent of the barrier height (as illustrated in \cref{figT06}{\color{blue}(a)}) due to pseudospin conservation. As a result, quasiparticles with small incident angles pass through the barrier with high probability, making it nearly transparent. In contrast, the minimum of the conductance depends on the barrier width; it decreases as the barrier width increases (as illustrated in \cref{figT06}{\color{blue}(b)}). Theoretically, the degeneracy of electron and hole states at the critical limit $E=U_{0}=0$ can lead to a vanishing minimum conductance. However, experimental observations show a finite minimum conductance at $E=U_{0}=0$ \citep{10.1063/1.2980045}.

\section{CONCLUSION}\label{tsection4}

In summary, we have investigated Klein tunneling and Fabry-Pérot resonances of charge carriers in twisted bilayer graphene through a rectangular potential barrier structure. We start our study by providing the effective Hamiltonian that describes our system, then we derived the corresponding energy bands. Using this Hamiltonian, we have numerically evaluated the transmission probabilities and conductance of quasiparticles at normal and non-normal incidence, for $K$ and $K_{\theta}$ cones, when impinging on the barrier.

According to our numerical analysis, the results revealed that the behavior of Klein tunneling and Fabry-Pérot resonances is highly dependent on the barrier parameters, particularly the height and width of the potential barrier. As the barrier height increases, the transmission of quasiparticles becomes increasingly anisotropic, causing a notable deflection of both Klein tunneling and Fabry-Pérot resonances away from normal incidence. This deflection results from to the chirality of Dirac fermions and the interplay between the two Dirac cones in TBG. In contrast, the system does not exhibit a deflection effect by increasing the barrier width; instead, the positions of Klein tunneling and Fabry-Pérot resonances shift progressively from normal to non-normal incidence. This shift is due to the multiple internal reflections within the barrier region, which influence the interference conditions governing the formation of resonances.

Furthermore, our analysis indicates that the conductance is significantly affected by the barrier height but shows only weak dependency on the barrier width. Specifically, our results shows that higher barrier heights decrease while simultaneously modifying the positions of conductance peaks, which are associated with Fabry-Pérot resonances. This tunability provides control over conductance maxima and minima, which may offer new possibilities for designing electronic devices that exploit anisotropic transport properties. These results suggest that by carefully adjusting the barrier parameters, it is possible to manipulate charge carrier dynamics in TBG, leading the way for novative applications in electronic nanodevices and quantum transport engineering.

These findings provide new insights into tunneling phenomena in TBG and highlight the potential for engineering electronic transport properties in 2D materials. By enabling control over transmission resonances and conductance, our work paves the way for applications in electronic nanodevices, electron optics, and valleytronics.

\bibliography{TBGreson}
\bibliographystyle{apsrev4-2}
\end{document}